\documentclass[aps,showpacs,preprint,epsf,epsfig, nofootinbib]{revtex4-1}
\usepackage{graphicx}
\usepackage{dcolumn}
\usepackage{bm}
\usepackage{hyperref}
\usepackage{epsfig,graphicx,color,appendix}
\usepackage{multirow}
\usepackage{tabularx}
\usepackage{capt-of}
\usepackage{amsmath}
\usepackage{caption}
\usepackage[countmax]{subfloat}

\usepackage{threeparttable}
\begin{document}
\def\ben{\begin{eqnarray}}
\def\en{\end{eqnarray}}
\def\non{\nonumber}
\def\la{\langle}
\def\ra{\rangle}
\def\t{\times}
\def\pp{{\prime\prime}}
\def\nc{N_c^{\rm eff}}
\def\vp{\varepsilon}
\def\hep{\hat{\varepsilon}}
\def\a{{\cal A}}
\def\B{{\cal B}}
\def\c{{\cal C}}
\def\d{{\cal D}}
\def\e{{\cal E}}
\def\p{{\cal P}}
\def\tt{{\cal T}}
\def\up{\uparrow}
\def\dw{\downarrow}
\def\vma{{_{V-A}}}
\def\vpa{{_{V+A}}}
\def\smp{{_{S-P}}}
\def\spp{{_{S+P}}}
\def\J{{J/\psi}}
\def\ov{\overline}
\def\Lqcd{{\Lambda_{\rm QCD}}}
\def\pr{{Phys. Rev.}~}
\def\prl{{ Phys. Rev. Lett.}~}
\def\pl{{ Phys. Lett.}~}
\def\np{{ Nucl. Phys.}~}
\def\zp{{ Z. Phys.}~}
\long\def\symbolfootnote[#1]#2{\begingroup%
\def\thefootnote{\fnsymbol{footnote}}\footnote[#1]{#2}\endgroup}
\def\lsim{ {\ \lower-1.2pt\vbox{\hbox{\rlap{$<$}\lower5pt\vbox{\hbox{$\sim$}
}}}\ } }
\def\gsim{ {\ \lower-1.2pt\vbox{\hbox{\rlap{$>$}\lower5pt\vbox{\hbox{$\sim$}
}}}\ } }

\font\el=cmbx10 scaled \magstep2{\obeylines\hfill \today}

\vskip 1.5 cm

\centerline{\large\bf First Estimates of Nonleptonic $\pmb B_c \to AT$ Weak Decays }

\small
\vskip 1.0 cm

\centerline{\bf Rohit Dhir \symbolfootnote[2]{dhir.rohit@gmail.com} and C. S. Kim \symbolfootnote[3]{cskim@yonsei.ac.kr} }
\medskip
\centerline{\it Department of Physics and IPAP, Yonsei University, Seoul 120-749, Korea}
\bigskip
\bigskip
\begin{center}
{\large \bf Abstract}
\end{center}
We obtain the first estimates of branching ratios for the weak decays of $B_c$ meson decaying to two orbitally excited mesons namely, axial-vector meson ($A$) and tensor meson ($T$), in the final state. We calculate $B_c \to T$ transition form factors using the Isgur-Scora-Grinstein-Wise II framework and consequently predict branching ratios of $ B_c \to AT $ decays in Cabibbo-Kobayashi-Maskawa favored and suppressed modes.

\medskip
\medskip

PACS Numbers: 12.39.St, 12.39.Jh, 13.25.Hw, 13.25.Jx, 14.40.Nd
\vfill

\section{Introduction}

The  $B_c$ meson was first discovered through the observation of semileptonic decay, $ B_c
\to J/\psi l^+ \nu X $  by the CDF collaboration \cite{1}. Later on, more precise measurements of its life time and mass are confirmed to be $(0.453 \pm 0.042) $ ps and $(6277 \pm 6) $ MeV, respectively \cite{2,3}. A few of $B_c$ decay channels have also been observed experimentally. A new era of $B_c$ has started with an expected production cross-section of $\sim 0.4 ~\mu$b at centre-of-mass energy $\sqrt{s}=7 $ TeV at the LHC \cite{4}. Very recently, the LHC-b has reported an observation of  decays like $B_c^+ \to \psi (2S) \pi^+ $, $B_c^+ \to J/\psi \pi^+ \pi^+ \pi^-$ and $B_c^+ \to J/\psi \pi^+$ \cite{3, 4}. As a result, investigation of $B_c$ meson has become one of the most interesting current topic. Moreover, the LCH-b is expected to produce $\mathcal{O}(10^{10})$ events per year \cite{3,4,5,6,7}, which would provide a rich amount of information regarding $B_c$ meson.

 $B_c$ meson is the Standard Model (SM) particle composed of two flavors of heavy quarks, charm ($c$) and beauty ($b$). The $B_c ~(b \bar{c})$  meson being flavor asymmetric behaves differently from the heavy quarkonia ($b \bar{b}$, $c\bar{c}$). It only decays \textit{via} weak interactions as compared to heavy quarkonium states which can decay \textit{via} strong interactions and/or electromagnetic interactions.
 The decay processes of the $B_c$ meson involve decay of any  constituent quarks \textit{b} and \textit{c} with other being a spectator.
They can also annihilate weakly to produce leptons or lighter mesons which, being relatively suppressed, are ignored in the present analysis. The two-body  weak decays of $B_c$ may provide a good scenario to understand QCD dynamics both in perturbative and nonperturbative regime, to test QCD-motivated theories/models and to study physics beyond the SM.

Hadronic two-body decays of $B_c$ mesons have broadly been studied in recent years. There exist comprehensive literature on phenomenological works based on different approaches to study semileptonic and nonleptonic decays of $B_c$ meson emitting $s$-wave mesons in the final state \cite{6,7,8,9,10,11,12,13,14,15,16,17,18}. Recently, the focus has been shifted towards the $p$-wave emitting decays of $B_c$ meson \cite{19,20,21,22,23,24,25,26,27,28,29}. In our previous work \cite{29}, we have studied the $B_c$ meson decaying to two $p$-wave in the final state \textit{i.e.} $B_c \to AA$ decays, and shown that branching ratios of these decays are comparable to $B_c \to VA$ modes. Also, the Particle Data Group (PDG) \cite{2} has reported many single $p$-wave meson emitting decays having branching ratios of $\mathcal{O} (10^{-6})$. This motivated us to carry forward our analysis to nonleptonic two-body $B_c$ decays considering that both mesons in final state are orbitally excited mesons (or $p$-wave mesons). Particularly, we study $B_c$ decaying to an axial vector ($A$) and tensor meson ($T$), which could compete with $B_c \to VT$ modes and their branching fractions could be measured in near future at the LHC-b experiment and at future $B$-factories. Moreover, $B_c \to AT$ decays can offer a scenario to study polarization of final state particles \cite{2}.
In the present work, we obtain the first estimates of branching ratios of exclusive $B_c \to AT$ decays. We calculate $B_c \to T$ form factors using  the improved Isgur-Scora-Grinstein-Wise quark model framework (called ISGW II Model) \cite{30,31}. The ISGW II model is one of the few successful models \cite{30,31,32,33} which is used to calculate $B_c \to T$ transition form factors.

This paper is organized as follows: In Section II, we discuss briefly about the effective weak Hamiltonian and decay width formula. We give input parameters like mixing scheme and decay constants for axial vector mesons, and $B_c \to T$ transition form factors in Section III.  Results and discussions are given in Sec. IV. In final section, we present summary and conclusions.

\section{Methodology}
\subsection{Weak Hamiltonian}

 The QCD modified weak Hamiltonian \cite{34} generating the $B_c$ decay involving $b\to c$ and $b \to u$ transitions in
 Cabibbo-Kobayashi-Maskawa (CKM) favored modes ($\Delta b = 1, \Delta C =1, \Delta S=0 ;~~\Delta b = 1, \Delta C =0,\Delta S =-1$)  and  CKM suppressed ($\Delta b = 1, \Delta C=1, \Delta S=-1; ~~\Delta b = 1, \Delta C=1, \Delta S=1; ~~\Delta b = 1, \Delta C=-1, \Delta S=-1; ~~\Delta b = 1, \Delta C=-1, \Delta S=0$) modes is expressed as follows:
\begin{eqnarray}
H_{w}^{(\Delta b=1)} &=& \frac{G_F}{\sqrt{2}}\displaystyle\sum\limits_{Q (q)=u,c}\displaystyle\sum\limits_{q^{'}=d,s}V^*_{Qb}V_{qq^{'}} \Big(a_1(\mu)
O^{qq^{'}}_1(\mu) + a_2(\mu) O^{qq^{'}}_2(\mu)\Big) + h.c.\ ;
\end{eqnarray}
 where $G_{F} $ is the Fermi constant and $V_{ij} $ are the CKM matrix elements. $a_{1} $ and $a_{2} $ are the standard perturbative QCD coefficients, evaluated at renormalization scale $\mu \approx \mathcal{O} (m_{b}) $. Local tree level operators $O_{1,2}$ involving $b \to q$ transition are given below:
  \begin{eqnarray}
  O^{qd}_1 &=& (\bar b_\alpha q_\alpha)_{V-A} \cdot (\bar q_\beta d_\beta)_{V-A},  \nonumber \\
O^{qd}_2 &=& (\bar b_\alpha q_\beta)_{V-A} \cdot (\bar q_\beta d_\alpha)_{V-A},  \nonumber \\
   O^{qs}_1 &=& (\bar b_\alpha q_\alpha)_{V-A} \cdot (\bar q_\beta s_\beta)_{V-A},  \nonumber \\
O^{qs}_2 &=& (\bar b_\alpha q_\beta)_{V-A} \cdot (\bar q_\beta s_\alpha)_{V-A},   
\end{eqnarray}
where $\bar{q}q' \equiv \bar{q} \gamma_{\mu}(1-\gamma_5)q'$, $\alpha$ and $\beta$ are $SU(3)$ color indices. In addition, $B_c$ meson can also decay via the bottom conserving modes, where the charm quark decays to an $s$ or a $d$ quark. However, such $B_c \to AT$ decays are kinematically forbidden.

By factorizing matrix elements of the four-quark operator contained in the effective weak Hamiltonian (1), one can divide these decays in three classes: Contribution from W-external emission diagram at tree level (color favored diagram) is classified as class I type decays. In this case the decay amplitudes are proportional to $a_{1} $, where $a_{1} (\mu )=c_{1} (\mu )+\frac{1}{N_{c} } \, c_{2} (\mu )$, and $N_{c} $ is the number of colors. Class II type decays consist of contribution from another W-internal emission diagrams (color suppressed diagrams). The decay amplitude in this class is proportional to $a_{2} (\mu )=c_{2} (\mu )+\frac{1}{N_{c} } \, c_{1} (\mu ).$  Class III transitions are caused by the interference of both color favored and color suppressed diagrams. Interestingly, we find that $B_c \to AT$ decays are independent of $a_1$ and $a_2$ interference.

For numerical calculations, we follow the $B$ physics convention of taking $N_{c} =3 $ to fix the QCD coefficients $a_{1} $ and $a_{2} $, \textit{i.e.}
$$a_{1} (\mu ) = 1.03 ,~a_{2} (\mu )= 0.11 $$
   where  we use \cite{34}:
     \begin{center}
 $c_{1} (\mu ) = 1.12 ,~c_{2} (\mu )=-0.26 $  at  $ \mu \approx m_{b}^{2}$.
\end{center}
We want to point out that $N_c$, the number of color degrees of freedom, is generally treated as a phenomenological parameter in weak meson decays to account for non-factorizable contributions \cite{34}. In the absence of any experimental and theoretical study of $B_c \to AT$, we also compare our results with those of large $N_c$ limit. The obtained results would provide a reasonable estimate of range of branching ratio between $N_c= 3$ to $N_c \to \infty$.

\subsection {Decay Amplitudes}

\par In the framework of generalized factorization the decay amplitudes $\langle AT|H_w|B_c \rangle$ are approximated as a product of the matrix elements of weak currents (up to the weak scale factor of $\frac{G_{F} }{\sqrt{2}} $ $\times$ CKM elements $\times$ QCD factor) given by
\ben
 \la TA  | H_{w}  | B_{c} \ra  \sim  \la  T  | J^{\mu}  | 0 \ra  \la  A  | J_{\mu }  | B_{c} \ra  & + & \la  A  | J^{\mu } | 0 \ra  \la  T | J_{\mu }  | B_{c}  \ra ,
 \en
where $J^{\mu} = V^{\mu} -A^{\mu}$, $V^{\mu}$ and $A^{\mu}$
denote a vector and an axial-vector current, respectively.
\par
In order to calculate the numerical values of decay amplitudes of $B_c \to AT$ decays, we have to obtain the above mentioned hadronic matrix elements.  Notice that the polarization tensor $\epsilon_{\mu\nu}$ of the $^{3}P_{2}$ tensor meson satisfies the following relations \cite{35}:
 \begin{equation}\label{}
\epsilon_{\mu\nu}(p_T,\lambda)=\epsilon_{\nu\mu}(p_T,\lambda), \qquad p_{\mu}\epsilon^{\mu\nu}(p_T,\lambda)=p_{\nu}\epsilon^{\mu\nu}(p_T,\lambda)=0, \qquad \epsilon_{\mu}^{\mu}(p_T,\lambda)=0,
\end{equation}
where $\lambda$ defines state of definite halicities. Consequently, the matrix element between the vacuum and final state tensor meson $T$ is
\begin{equation}\label{}
\langle 0|(V-A)_{\mu}|T\rangle = a\epsilon_{\mu\nu}p^{\nu}(p_T,\lambda)+ b \epsilon^{\nu}_{\nu}p_{\mu}(p_T,\lambda)=0.
\end{equation}
This relation in general follows from Lorentz covariance and parity consideration. Hence the tensor meson cannot be produced from the $V-A$ current; \textit{i.e.}, the decay constant of the tensor meson vanishes. This fact further simplifies the decay amplitude $\la AT  | H_{w}  | B_{c} \ra$ expressed by relation (3), which yields
\ben
\left\langle {AT} \right|{H_w}\left| {{B_c}} \right\rangle  \sim \la  A  | J^{\mu } | 0 \ra  \la  T | J_{\mu }  | B_{c}  \ra .
\en
On the other hand, matrix element between the vacuum and final state axial-vector meson is defined as
\ben
\la A(p_{A} ,\vp ) | A_{\mu } | 0 \ra  &=& \vp _{\mu }^{*} m_{A} f_{A} ,
\en
where $\vp _{\mu }$ denotes polarization of axial-vector meson and $f_A$ is the corresponding decay constant.

 Using Lorentz invariance, the hadronic transition matrix elements \cite{30,31} for the relevant weak current between meson states can be parameterized as follows:
\begin{eqnarray}
  \langle T|V^{\mu}|B_c\rangle &=& i h(q^{2}) \varepsilon^{\mu\nu\rho\sigma} \epsilon_{\nu\alpha}p_{B_c}^{\alpha} (p_{B_c}+p_{T})_{\rho}(p_{B_c}-p_{T})_{\sigma}, \nonumber\\
  \langle T|A^{\mu}|B_c\rangle &=& k(q^{2}) \epsilon^{*\mu\nu} (p_{B_c})_{\nu}+ \epsilon_{\alpha\beta}^{*}p_{B_c}^{\alpha}p_{B_c}^{\beta} \nonumber\\
  & & \left[ b_{+}(q^{2})(p_{B_c}+p_{T})^{\mu} + b_{-}(q^{2})(p_{B_c}-p_{T})^{\mu} \right],
\end{eqnarray}
where $p_{B_c}$ and $p_T$ denote the
momentum of the $B_c$ meson and the tensor meson $T$, respectively such that $q^{\mu} \equiv p_{B_c}^{\mu} -p_T^{\mu}$. $h(q^2)$, $k(q^2)$, $b_+(q^2)$, and $b_-(q^2)$ represent the relevent
form factors for the $B_c \rightarrow T$ transition,
$F^{B \rightarrow T}(m_{B_c}^2)$, which have been
calculated at $q^2 = q^2_{max}$
using the ISGW II quark model \cite{31}.

Using definitions (7) and (8), the matrix element (6) takes the form
\ben
A({B_c}\, \to \,A\,T) = \frac{{{G_F}}}{{\sqrt 2 }} \times ({\rm CKM\,factors \times QCD\,factors}) \times {f_A} m_A{ \in ^{*\alpha \beta }}F_{\alpha \beta }^{{B_c} \to T}(m_A^2),
\en
where
  \ben
F_{\alpha \beta }^{{B_c} \to T} =  \epsilon _\mu ^*{({p_{{B_c}}} + {p_T})_\rho }[ih{\varepsilon ^{\mu \nu \rho \sigma }}{g_{\alpha \nu }}{({p_A})_\beta }{({p_A})_\sigma } + k\delta _\alpha ^\mu \delta _\beta ^\rho  + {b_ + }{({p_A})_\alpha }{({p_A})_\beta }{g^{\mu \rho }}].
\en

\subsection{Decay Widths}
The simplified decay width formula, after summing over polarizations of the tensor meson $T$,  for the $B_c \to  A T$ decays is given as
\begin{align}\label{AT}
\Gamma(B_c \to  A T) =& \frac{G_{F}^{2}}{6\pi m_{T}^{4}} m_{B_c}^2 f_{A}^{2} \times \left[ \alpha(m^{2}_{A})|\bm{p_A}|^{7} + \beta(m^{2}_{A})|\bm{p_A}|^{5}+ \gamma(m^{2}_{A})|\bm{p_A}|^{3} \right],
\end{align}

\noindent where $|\bm{p_A}| (=|\bm{p_T}|)$ is the magnitude of three momentum of the final state particle in the rest frame of $B_c$ meson. $|\bm{p_A}|$ can be expressed as $|\bm{p_A}|= \sqrt{\lambda}/2m_{B_c}$, where  $\lambda \equiv \lambda(m^{2}_{B},m^{2}_{T},m^{2}_{A})= (m^{2}_{B_c}+m^{2}_{T}-m^{2}_{A})^{2} - 4 m^{2}_{A}m^{2}_{T}$ is the triangle function. The coefficients $\alpha$, $\beta$
and $\gamma$ are quadratic functions of  form factors $k$, $b_{+}$ and $h$, evaluated at $q^2 = m^{2}_{A}$. These coefficients are given as follows:

\begin{equation}
\alpha(m^{2}_{A}) = 8 b_{+}^{2} m_{B_c}^{2},
\end{equation}

\begin{equation}
\beta(m^{2}_{A}) = \frac{1}{4} \left[k^{2}+6m_{T}^{2}m^{2}_{A}h^{2}+2(m_{B}^{2}-m_{T}^{2}-m^{2}_{A})kb_{+}\right],
\end{equation}

\begin{equation}
\gamma(m^{2}_{A}) = \frac{5 k^{2} m^{2}_{A} m_{T}^{2}}{8 m_{B_c}^2}.\\
\end{equation}

\section{Input Parameters}
\subsection{Mixing Angles for Axial-Vector Mesons}
Mixing scheme of the axial-vector mesons have thoroughly been discussed in the literatures \cite{32,33,36,37,38,39,40,41,42}. Thus, we briefly mention the important facts. There are two types of axial-vector mesons, in spectroscopic notation ${}^{2s+1}L_J$, ${}^{3} P_{1} $($J^{PC} =1^{++} $) and ${}^{1} P_{1} $ $(J^{PC} =1^{+-} )$, respectively. These states can mix in two ways: first, states ${}^{3} P_{1} $ or ${}^{1} P_{1} $ can mix within themselves; second, mixing between ${}^{3} P_{1} $ or ${}^{1} P_{1} $ states.  Experimentally observed non-strange and uncharmed ${}^{3} P_{1} $ meson sixteenplet consists of isovector $a_{1}(1.230) $  and four isoscalars $f_{1} (1.285)$, $f_{1} (1.420)$/$f'_{1} (1.512) $ and $\chi _{c1} (3.511)$. On the other hand, ${}^{1} P_{1} $ meson multiplet includes isovector $b_{1} (1.229)$ and three isoscalars $ h_{1} (1.170)$, $ h'_{1} (1.380)$ and $h_{c1} (3.526)$, where $h_{c1} (3.526)$ and $ h'_{1} (1.380)$ are not well understood experimentally\footnote{ 
Here the quantities in brackets indicate their respective masses (in GeV).}.

The following mixing scheme have been proposed for the isoscalar ($1^{++} $)and ($1^{+-} $) mesons:
\ben
 f_{1} (1.285) &=& \frac{1}{\sqrt{2} } (u\overline{u}+d\overline{d})\cos \phi_{A} +(s\overline{s})\sin \phi_{A} \non
  \\ f'_{1} (1.512)\,&=&\frac{1}{\sqrt{2} } (u\overline{u}+d\overline{d})\sin \phi_{A} -(s\overline{s})\cos \phi _{A} \non
 \\ \chi _{c1} (3.511)&=&(c\bar{c}),
\en
and
\ben
 h_{1} (1.170)&=&\frac{1}{\sqrt{2} } (u\overline{u}+d\overline{d})\cos \phi {}_{A'} +(s\overline{s})\sin \phi {}_{A'}, \non \\  h'_{1} (1.380)&=&\frac{1}{\sqrt{2} } (u\overline{u}+d\overline{d})\sin \phi _{A'} -(s\overline{s})\cos \phi _{A'} , \non
\\h_{c1} (3.526)&=&(c\bar{c}),
\en
respectively, with

\[\phi _{A(A')} =  \theta (ideal) - \theta _{A(A')} (physical).\]

The observation that $f_{1}(1.285) \to 4\pi $/$\eta \pi \pi $, $f_{1}^{'} (1.512) \to K\bar{K}\pi $, $h_{1} (1.170) \to \rho \pi $ and $h_{1}^{'} \to K\bar{K}^{*} $/$\bar{K}K^{*} $ predominantly seems to favor the ideal mixing for these nonets $i.e.$, $\phi _{A}  =\phi _{A'}  = 0^{\circ }$.
The  hidden-flavor neutral $a_{1}(1.230) $  and $b_{1} (1.229)$ states cannot mix due to opposite C-parity. Also, the opposite G parities under isospin symmetry prevent mixing of charged $a_{1}(1.230) $  and $b_{1} (1.229)$ states. In contrast, states involving strange partners namely,  $K_{1A} $ and $K_{1A'} $ of $A\, (1^{++} )$ and $A'(1^{+-} )$ mesons, respectively, mix to generate the physical states in the following convention:
\ben
 K_{1} (1.270)&=& K_{1A}  \sin \theta _{K_1}  + K_{1A'}  \cos \theta _{K_1} ,\non
  \\ \underline{K}_{1} (1.400)&=& K_{1A}  \cos \theta _{K_1}  - K_{1A'}  \sin \theta _{K_1} .
\en

  Several phenomenological analyses indicate that the mixing angle $\theta_{K_1}$ lies in the vicinity of $\sim 35^{\circ} $ and $\sim 55^{\circ}$, see for details \cite{38}. Several studies based on the experimental information obtained twofold ambiguous solutions $\theta _{K_1} =\pm \, 37^{\circ} $ and $\theta _{K_1} =\pm \, 58^{\circ} $ \cite{36,37}. We wish to point out that the study of $D\to K_{1} (1.270) \pi$ /$ K_{1} (1.400) \pi $ decays and experimental measurement of the ratio of $K_1 \gamma$ production in $B$ decays favor negative mixing angle solutions \cite{32,33,37,41}. In a recent study \cite{38}, it has been argued that choice of angle for $f_1-f_1^{'}$ and $h_1-h_1^{'}$ mixing schemes is intimately related to choice of mixing angle $\theta_{K_1}$. The mixing angle $\theta_{K_1} \sim 35^{\circ}$ is preferred over $\sim 55^{\circ}$ for nearly ideal mixing for $f_1-f_1^{'}$ and $h_1-h_1^{'}$. However, we also give results at $\theta_{K_1}= -58 ^{\circ} $ in our numerical calculations.

Following the Heavy Quark Spin (HQS) scheme, the heavy axial-vector resonances (likely charmed and bottom axial vector meson states) are generally taken to be the mixture of $ P ^{1 /2}_1$ and $P ^{3 /2}_1 $ states. In the heavy quark limit, the heavy quark spin $S_{Q} $ and the total angular momentum of the light antiquark can be used as good quantum numbers, separately \cite{37,40}. Therefore, in heavy axial vector resonances, the physical mass eigenstates $P_{1}^{3/ 2}  $ and $P_{1}^{1/2} $ with $J^{P} =1^{+} $ can be expressed as a mixture of ${}^{3} P_{1} $ and ${}^{1} P_{1} $ states \textit{i.e.}
\ben
 |P_{1}^{1/2 } > &=& -\sqrt{\frac{1}{3} } |{}^{1} P_{1} > + \sqrt{\frac{2}{3} } |{}^{3} P_{1} >,\non \\ |P_{1}^{3/2} >&=& \sqrt{\frac{2}{3} } |{}^{1} P_{1} >+\sqrt{\frac{1}{3} } |{}^{3} P_{1} >.
 \en
 In the heavy quark limit, the physical states $D_{1} (2.427)$ and $\underline{D}_{1} (2.422)$ can be identified as $P_{1}^{1/2} $ and $P_{1}^{3/2} $, respectively. However, beyond the heavy quark limit, there can be mixing between $P_{1}^{1/2} $ and $P_{1}^{3/2} $ given as
\ben
D_{1} (2.427)&=& D_{1}^{1/2} \cos \theta _{D_{1}}  + D_{1}^{3/2}  \sin \theta _{D_{1}} , \non
\\
\underline{D}_{1} (2.422)&=& -D_{1}^{1/2}  \sin \theta _{D_{1}}  + D_{1}^{3/2}  \cos \theta _{D_{1}} .
\en
Likewise, mixing scheme for strange charmed axial-vector mesons is expressed as,
\ben D_{s1} (2.460)&=& D_{s1}^{1/2} \cos \theta _{D_{s1}}  + D_{s1}^{3/2}  \sin \theta _{D_{s1}} ,\non \\ \underline{D}_{s1} (2.535)&=& -D_{s1}^{1/2 }  \sin \theta _{D_{s1}3}  +D_{s1}^{3/2 } \cos \theta _{D_{s1}} .
\en
Following the analysis given by the Belle \cite{39}, we use the mixing angle $\theta _{D_{1}} =(-5.7\pm 2.4)^{\circ } $, while the quark potential model analysis \cite{40} yields $\theta _{D_{s1}} \approx 7^{\circ } $.

\subsection{Decay Constants}

The decay constants for tensor mesons vanish corresponding to the condition (5), while the decay constants for axial-vector mesons are defined by the reduced matrix elements expressed in relation (7) in the previous section. It is a well established fact that the axial-vector meson ${}^{3} P_{1} $ and ${}^{1} P_{1} $ states transform under the charge conjugation \cite{30} as  \ben
 M_a^b(^3P_1) \to M_b^a(^3P_1), \qquad M_a^b(^1P_1) \to
 -M_b^a(^1P_1),~~~(a=1,2,3).
 \en
 Since the weak axial-vector current transfers as
$(A_\mu)_a^b\to (A_\mu)_b^a$ under charge conjugation, the decay constants of the $^1P_1$ mesons state vanish in the flavor symmetry limit \cite{30}. However, in the presence of symmetry breaking $^1P_1$ mesons may acquire non-zero values of the decay constants. In case of non-strange axial-vector mesons, $f_{a1}= 0.203 \pm 0.018 $ GeV is quoted in the analysis given by Bloch \textit{et. al} \cite{42}. The
value $f_{a1}= 0.238 \pm 0.010 $ GeV is obtained using the QCD sum rule method \cite{43}. Nardulli and Pham \cite{41} used SU(3) symmetry to determine $f_{a_{1} } =0.223 (-0.215)$ GeV for $\theta_{K_1} = -58^\circ (-32^\circ)$ mixing angle for strange axial vector mesons. Since, $a_1$ and $f_1$ lies in the same $SU(3)$ nonet we assume $f_{f_{1} } \approx f_{a_{1} } $. In isospin limit, owing to G-parity conservation decay constant $f_{b_{1}}= 0$. Experimental data on $\tau \to K_{1}(1270) \nu_{\tau}$ decay yields the decay constant $f_{K_{1}} (1270) =0.175 \pm 0.019$ GeV \cite{20,31}, while decay constant for $\underline{K}_{1} (1.400) $ may be calculated by using relation $f_{\underline{K}_{1}} (1.400)/f_{K_{1}} (1.270) \approx \cot \theta _{K_1} $ $i.e.$
$f_{\underline{K}_{1}} (1.400) =(-0.109\pm 0.012)$ GeV, for  $\theta _{K_1} =-58^{\circ}$ ; $f_{\underline{K}_{1}} (1.400) =(-0.232\pm 0.025)$ GeV, for $\theta _{K_1} =-37^{\circ}$;
 used in the present work \cite{31}.
For decay constants of axial vector charmed and strange charmed meson states, we have used
$f_{D_{1A}} =-0.127$ GeV, $f_{D_{1A}^{'} } =0.045$ GeV, $f_{D_{s1A}} =-0.121$ GeV, $f_{D_{s1A}^{'} } =0.038$ GeV, and $f_{\chi _{c1} } \approx -0.160$ GeV  for numerical evaluation \cite{29,32,33,37}. It may be noted that the decay constants of $^3P_1$ states have opposite signs to that of $^1P_1$ as apear from (21).

\subsection{Form factors}
In Isgur-Scora-Grinstein-Wise model \cite{31}, weak hadronic transition form factors are predicted using non-relativistic quark model wave functions.
The form factors obtained are assumed to be reliable at near the zero recoil point where $ q^2$ reaches its maximum value $(m_B-m_X)^2$. The problem being that the form-factor $q^2$-dependence in the original ISGW model \cite{30} is proportional to  $e^{ -(q^2_m-q^2)}$ as a result form factor decreases exponentially. Nevertheless, this has been improved in updated version of ISGW model (ISGW II) \cite{30,31} where the form factors show a more realistic behavior at large $(q^2_m-q^2)$. The exponential dependence of $B_c \to T$ form factors has been replaced by a polynomial term. In addition,  ISGW II model incorporates heavy quark symmetry constraints, heavy-quark-symmetry-breaking color magnetic interaction, relativistic corrections, and \textit{etc.}

The simplified expressions for $h$, $k$, $b_+$ and $b_-$ form factors in the ISGW II model for $B_{c} \to T$ transitions are given as \cite{30,31}:

\ben
h & = &\frac{{{m_d}}}{{2\sqrt {2{{\tilde m}_{{B_c}}}{\beta _{{B_c}}}} }}\,\left( {\frac{1}{{{m_q}}} - \frac{{{m_d}\beta _{{B_c}}^2}}{{2{\mu _ - }{{\tilde m}_T}\beta _{{B_c}T}^2}}} \right)F_5^{(h)},\en

\ben
k & = & \frac{{{m_d}}}{{\sqrt {2{\beta _{{B_c}}}} }}\,(1 + \tilde \omega )\,\,F_5^{(k)},
\en
\ben
{b_ + } + {b_ - } & = &\frac{{m_d^2}}{{4\sqrt 2 {m_q}{m_b}{{\tilde m}_{{B_c}}}{\beta _{{B_c}}}}}\,\frac{{\beta _T^2}}{{\beta _{{B_c}T}^2}}\left( {1 - \frac{{{m_d}}}{{2{{\tilde m}_{{B_c}}}}}\frac{{\beta _T^2}}{{\beta _{{B_c}T}^2}}} \right)\,\,F_5^{({b_ + } + {b_ - })},
\en
\ben
{b_ + } - {b_ - } & = &  - \frac{{{m_d}}}{{\sqrt 2 {m_b}{{\tilde m}_T}{\beta _{{B_c}}}}} F_5^{({b_ + } - {b_ - })} \non \\  & ~ & \t \left( 1 - \frac{{m_d}{m_b}}{2{\mu _ + }{\tilde m}_{B_c}}\frac{{\beta _T^2}}{{\beta _{{B_c}T}^2}} + \frac{{{m_d}}}{{4{m_q}}}\frac{{\beta _T^2}}{{\beta _{{B_c}T}^2}} \left( {1 - \frac{{{m_d}}}{{2{{\tilde m}_{{B_c}}}}}\frac{{\beta _T^2}}{{\beta _{{B_c}T}^2}}} \right) \right),
\en
where
\ben
\mu _{\pm } =(\frac{1}{m_{c} } +\frac{1}{m_{b} } )^{-1} .
\en
 $t(\equiv q^{2} )$ dependence is given by
\ben
\tilde{\omega }-1=\frac{t_{m} -t}{2\bar{m}_{B_{c} } \bar{m}_{A} },
\en
where $t_{m} =(m_{B_{c} } -m_{T} )^{2} $ is the maximum momentum transfer,
and
\ben
 F_{5}^{(h)} &=& F_{5} (\frac{\bar{m}_{B_{c} } }{\tilde{m}_{B_{c} } } )^{-3/2} (\frac{\bar{m}_{T} }{\tilde{m}_{T} } )^{-1/2} , \non \\ F_{5}^{(k)} &=&  F_{5} (\frac{\bar{m}_{B_{c} } }{\tilde{m}_{B_{c} } } )^{-1/2 } (\frac{\bar{m}_{T} }{\tilde{m}_{T} } )^{1/2} , \non \\ F_{5}^{(b_{+} +b_{-} )} &=& F_{5} (\frac{\bar{m}_{B_{c} } }{\tilde{m}_{B_{c} } } )^{-5/2} (\frac{\bar{m}_{T} }{\tilde{m}_{T} } )^{1/2 }, \non \\
  F_{5}^{(b_{+} -b_{-})} &=& F_{5} (\frac{\bar{m}_{B_{c} } }{\tilde{m}_{B_{c} } } )^{-3/2} (\frac{\bar{m}_{T} }{\tilde{m}_{T} } )^{-1/2},
\en
where $\tilde{m}$ represents sum of the meson constituent quarks masses, $\bar{m}$ is the hyperfine averaged physical masses. The parameter $\beta $ have different values for \textit{s}-wave and \textit{p}-wave mesons as shown in the Table I \cite{26,27}.

The function $F_5$ is given by
\ben
F_{5} = \left(\frac{\tilde{m}_{T} }{\tilde{m}_{B_{c} } } \right)^{1/2 } \left(\frac{\beta _{B_{c} } \beta _{T} }{\beta _{B_{c} T} } \right)^{5/2 } \left[1+\frac{1}{18} \chi ^{2} (t_{m} -t)\right]^{-3} ,
\en
where
\ben
\chi ^{2}  = \frac{3}{4m_{b} m_{c} } +\frac{3m_{c}^{2} }{2\bar{m}_{B_{c} } \bar{m}_{T} \beta _{B_{c} T}^{2} } +\frac{1}{\bar{m}_{B_{c} } \bar{m}_{T} } \left(\frac{16}{33-2n_{f} } \right)\ln [\frac{\alpha _{S} (\mu _{QM} )}{\alpha _{S} (m_{c} )} ],
\en
and
\ben
\beta _{B_{c} T}^{2} =\frac{1}{2} \, \, \left(\beta _{B_{c} }^{2} +\beta _{T}^{2} \right).
\en
  $n_f = 5$, the number of active flavors and $\mu _{QM} $ is the quark model scale.  Following quark mass values (in GeV) have been used to evaluate $B_{c} \to T$ form factors:
 $$ m_u = m_d = 0.31,~ m_s = 0.49, ~m_c = 1.7,~  \textrm {and} ~ m_b = 5.0.$$

Finally the form factors thus obtained are given in Tables II.

\section{Results and Discussions}

In this section, we present branching ratios of nonleptonic $B_c \to AT$ decays using ISGW II model framework for CKM favored and CKM suppressed modes. The Branching ratios for $B_c$ decaying to an axial-vector meson and a tensor in the final state for CKM enhanced and CKM suppressed modes are given in column 3 of Tables III-V, respectively. We also present our results at $N_c \to \infty$ (see column 4 of Tables III-V) for the sake of comparison, as no theoretical or experimental information is available now. We observe the following:

\subsubsection{For CKM enhanced modes}
\begin{enumerate}
\item The most dominant decay channel in CKM enhanced bottom changing and charm conserving ($\Delta S=-1$) mode is $B_c$ meson decaying to $p$-wave charmonium $\chi_{c2}$  in the final state \textit{i.e.} Br($B_c^- \to a_1^- \chi_{c2} $) = $4.69 \t 10^{-4}$. The branching ratio increases by 15 $\%$ larger at large $N_c$ limit \textit{i.e.} $ 5.54 \t 10^{-4}$. However, our prediction is roughly half of the numerical value obtained by Chang \textit{et al.} \cite{44}, which is based on the Bethe-Salpeter equation and QCD inspired potential approach at $N_c \to \infty$. The next order branching ratio are of $B_c^- \to D_1^0 D_2^-$ and $B_c^- \to \underline{D}_1^0 D_2^-$ decays. It may be noted that at large $N_c$ limit, branching ratios for color favored modes increase roughly by 15$\%$, while that of color suppressed modes increase by $\sim 80 \%$ due to change in Wilson coefficients $a_1$ and $a_2$, respectively. However, our results remain unaffected from the interference of $a_1$ and $a_2$ terms as class III decays are not possible for $B_c \to AT$ decays.

\item In CKM enhanced bottom and charm changing ($\Delta S=0$) mode, the highest order branching ratios is: Br($B_c^- \to D_{s1}^{-} \chi_{c2}$) = $3.34 \t 10^{-5} ~(3.95 \t 10^{-5})$. Next order values are: Br($B_c^- \to \underline{D}_{s1}^{-}  \chi_{c2} $) = $7.49 \t 10^{-6} ~(8.85 \t 10^{-6})$ and Br($B_c^- \to \chi_{c1} D_{s2}^{-} $) = $4.78 \t 10^{-7}~(2.67 \t 10^{-6})$, where the numbers in the parenthesis are BRs at large $N_c$ limit. The  remaining decay modes are suppressed having branching ratios of $\mathcal{O} (10^{-8} \sim 10^{-10})$.

\item We wish to point out that the branching ratios of $B_c \to A(^3 P_1)T$ decays are larger than  $B_c \to A(^1 P_1)T$ decays due to larger values of respective decay constants. However, this trend is reversed in decays involving strange axial vector mesons for both angles \textit{i.e.} $\theta_{K_{1}}=-37^\circ $ or $ -58^\circ$ .
\item It may also be noted that change in mixing angle $\theta_{K_{1}}$ from $-37^\circ $ to $ -58^\circ$ does not affect the branching ratios of decays involving $K_1 (^3P_1)$, whereas branching ratios of decays involving $\underline{K_{1}} (^1P_1)$ are almost doubled.

\end{enumerate}
\subsubsection{For CKM suppressed and doubly suppressed modes}
\begin{enumerate}
\item In CKM suppressed bottom and charm changing mode ($\Delta S=-1$), branching ratios of dominant decays are  Br($B_c^- \to \underline{K}_{1}^{-} \chi_{c2}$) = $2.40 \t 10^{-5} ~(2.84 \t 10^{-5})$ and Br($B_c^- \to  K_{1}^{-} \chi_{c2})$ = $1.66 \t 10^{-5} ~(1.97\t 10^{-5})$ for ${\theta_{K_{1}}=-37^\circ}$; here also, numerical values in brackets indicate BRs at $N_c=\infty$. It is seen that branching ratio for $B_c^- \to \underline{K}_{1}^{-} \chi_{c2}$ decay increases by a factor of 2 for ${\theta_{K_{1}}=-58^\circ}$. The branching ratios of the remaining decay modes are of $\mathcal{O} (10^{-7} \sim 10^{-8})$.
\item As compared to ($\Delta C= 1, ~\Delta S=-1$), the dominant decay channels in ($\Delta C = \Delta S= 0$) have branching ratios of $\mathcal{O} (10^{-6}\sim 10^{-7})$ \textit{i.e.} Br($B_c^- \to D_1^{-} \chi _{c2}$) = $2.32 \t 10^{-6}~ (2.74 \t 10^{-6})$; Br($B_c^- \to \underline{D}_1^{-} \chi _{c2}$) = $7.42 \t 10^{-7} ~(8.77\t 10^{-7})$ and Br($B_c^- \to \bar{D}_2^{0} a_1^-$) = $6.82 \t 10^{-7} ~(8.07 \t 10^{-7})$. However, rest of the decays have branching ratios of $\mathcal{O} (10^{-8} \sim 10^{-12} )$.
\item As expected, the branching ratios of CKM suppressed decays are an order smaller than CKM enhanced modes both for strangeness changing and strangeness conserving channels.

\item Decay channels in Cabibbo doubly suppressed ($\Delta C= -1, ~\Delta S=-1$) and  ($\Delta C= -1, ~\Delta S=0$) modes remain highly suppressed except for Br($B_c^- \to  \underline{D}_{s1}^{-}\bar{D}_{2}^0$) $= 2.53 \t 10^{-7} ~(3.00 \t 10^{-7})$. Other decays in these channels have branching ratios $\mathcal{O}(10^{-8} \sim 10^ {-10})$.
\item Here also, the branching ratios of decays involving $ A(^3 P_1)$ in the final state are larger in comparison to the  decays involving $A(^1 P_1)$ except for strange axial vector mesons.
\end{enumerate}

In addition, some decay channels are also possible through annihilation contributions but are ignored in our analysis, as these are suppressed due to helicity and color arguments.

As a test of factorization approximation, we propose to study the ratio of branching fractions of $B_c \to AT$ to $B_c \to VT$ decays for the same quark content \textit{i.e.}
\begin{equation}\label{ratio}
\frac{Br(B_c \to AT)}{Br(B_c \to VT)} = \bigg( \frac{f_A}{f_V} \bigg)^{2} \left[\frac{ \alpha(m^{2}_{A})~|\bm{p_{A}}|^{7} + \beta(m^{2}_{A})~|\bm{p_{A}}|^{5}+ \gamma(m^{2}_{A})~|\bm{p_{A}}|^{3}}{ \alpha(m^{2}_{V})~|\bm{p_{V}}|^{7} + \beta(m^{2}_{V})~|\bm{p_{V}}|^{5}+ \gamma(m^{2}_{V})~|\bm{p_{V}}|^{3}} \right],
\end{equation}
which is independent of QCD coefficients, kinematic constants and CKM factors. We find this ratio for dominant modes in CKM enhanced and CKM suppressed modes as
$$ \frac{Br(B_c \to a_1^- \chi_{c2})}{Br(B_c \to \rho^- \chi_{c2})} \sim 0.9  ; ~~~~ \frac{Br(B_c \to \underline{K}_1^- \chi_{c2})}{Br(B_c \to K^{*-} \chi_{c2})}\Bigg|_{37^\circ} \sim 0.9 $$ and $$ \frac{Br(B_c \to K_1^- \chi_{c2})}{Br(B_c \to K^{*-} \chi_{c2})}\Bigg|_{37^\circ} \sim 0.6. $$
 It is clear from the expression (32) that this ratio may be used as a test of ISGW II model to determine $q^2$-dependence of the form factors provided that decay constants are known. Also, if the form factors are well known, relation (32) may be used to estimate the extent of mixing between $K_{1A}$ and $K_{1A'}$ states and consequently to determine $f_{K_{1A}}$ and $f_{K_{1A'}}$ decay constants.

With remarkable technological improvements in experiments and availability of high precision instrumentation, branching ratios of the order of $(10^{-6})$ could be measured precisely \cite{7} at the LHC, the LHC-b and Super-B factories in near future, which can provide the necessary insight in to the phenomenological studies related to $B_c$ meson physics.

\section{Summary and Conclusions}

In the present work, we have investigated $B_c \to A T$ decays in ISGW II model framework. We have obtained their branching ratios of $B_c \to T$ transition form factors in this model. We have presented results both at $N_c=3$ and large $N_c$ limit. We conclude the following:
\begin{enumerate}
\item In CKM enhanced ($\Delta b= 1,~\Delta C= 1, ~\Delta S=0$) the only dominant decay $B_c^- \to a_1^- \chi_{c2}$ has branching ratio ($2.32 \t 10^{-4}$). However, the dominant decays  $B_c^- \to D_{s1}^{-} \chi_{c2}$ and $B_c^- \to  \underline{D}_{s1}^{-}\chi_{c2}$ in ($\Delta b= 1,~\Delta C= 0, ~\Delta S=-1$) have branching ratios around $10^{-5}$ and $10^{-6}$, respectively. Branching ratios of all the decay modes range from $ (10^{-4}\sim 10^{-12}) $.
\item In CKM suppressed modes, the branching ratios are further suppressed by an order of magnitude as compared to CKM enhanced modes. Branching ratios for the dominant decays $B_c^- \to K_{1}^{-} \chi_{c2}$, $B_c^- \to \underline{K}_{1}^{-} \chi_{c2}$ and $B_c^- \to D_{1}^{-} \chi_{c2} $ are of $\mathcal{O}$ ($10^{-5} \sim 10^{-6}$).
\item In general, branching ratios of $B_c \to A(^3P_1)T$ decays are higher in comparison to the $B_c \to A(^1P_1)T$ decays due to larger decay constant of $A(^3P_1)$ mesons. However, this is reversed for decays involving $K_1$-meson in the final state.

\item Change in mixing angle $\theta_{K_{1}}$ from $-37^\circ $ to $ -58^\circ$ increases branching ratios of decays involving $\underline{K_{1}} (^1P_1)$ by a factor of two, however decays involving $K_1 (^3P_1)$ remain almost unaffected.

\item At large $N_c$ limit, branching ratios of all the decay modes are enhanced due to increased   values of $a_1$ and $a_2$. Also, our results are independent of sign of $a_1$ and $a_2$ as Class III type of decays are not possible in $B_c \to AT$ mode.
\end{enumerate}

Recently, several p-wave meson emitting decays have been reported in the PDG \cite{2} whose branching ratios are $\mathcal{O}(10^{-6})$. Therefore, we hope that the predicted BRs would be measured soon as experiments like the LHC, the LHC-b and the KEK-B are expected to accumulate more than $10^{10}$ $B_c$ events per year.

\begin{acknowledgments}
The authors would like to thank R.C. Verma for useful comments and discussions. The work was supported by the National Research Foundation of Korea (NRF)
grant funded by Korea government of the Ministry of Education, Science and
Technology (MEST) (No. 2011-0017430) and (No. 2011-0020333).
\end{acknowledgments}

\newpage

\begin{table}
\captionof{table} {The values of $\beta$ parameter  for $s$-wave and $p$-wave mesons in the ISGW II quark model.}
\label{t1}
\begin{tabular}{ c c c c c c c c c c } \toprule
Quark content  & $u\bar{d}$ & $u\bar{s}$ & $s\bar{s}$ & $c\bar{u}$ & $c\bar{s}$ & $u\bar{b}$ & $s\bar{b}$ & $c\bar{c}$ & $b\bar{c}$   \\ \hline
$\beta _{s} $(GeV)  & 0.41  & 0.44  & 0.53 & 0.45  & 0.56 & 0.43  & 0.54 & 0.88 & 0.92 \\ \hline
$\beta _{p} $ (GeV)  & 0.28  & 0.30  & 0.33 & 0.33  & 0.38 & 0.35  & 0.41 & 0.52 & 0.60 \\ \hline
\end{tabular}
\end{table}

\begin{table}
\captionof{table} {$B_{c} \to T$ transition form factors at $q^2 _{max.} $ in the ISGW II quark model.}
\label{t2}
\begin{tabular}{ c c c c c c }
\toprule
Modes & Transition & $h$ & $k$ & $b_+$ & $b_-$ \\
\hline
\multirow {2}{*}{$\Delta b =1, \Delta C = 0, \Delta S = -1$} & $B_{c} \to D_{2}$  &$ 0.016 $ & $0.515 $ & $-0.008 $ & $0.010$   \\ \cline{2-6}
  & $B_{c} \to D_{s2}$ &$0.019$ & $0.684 $ &  $-0.010 $ & $0.013 $  \\ \hline
$ \Delta b =1, \Delta C = 1, \Delta S = 0$ & $B_{c} \to \chi_{c2}(c \bar {c})$ &$0.021$ & $1.306 $ & $-0.015 $ & $0.018$  \\
\hline
\end{tabular}
\end{table}

\begin{table}
\centering

\captionof{table} {Branching ratios of $B_{c}   \to   AT$ decays for CKM-enhanced modes.}
\label{t3}
\begin{tabular}{c c l l} \hline \hline

\multirow{2}{*}{Mode}  &  \multirow{2}{*}{Decays} & \multicolumn{2}{c}{Branching Ratios} \\ \cline{3-4}
 \multicolumn{2}{c} {}& $N_c =  3$ & $N_c \to  \infty$ \\  \hline
\multirow{4}{*}{$\Delta S =  -1$}
& $B_c^- \to a_1^-  \chi _{c2}$  &  $4.69 \t 10^{-4}$  & $5.54 \t 10^{-4}$ ($ 9.17 \t 10^{-4}$)\cite{44}\\ \cline{2-4}

& $B_c^- \to b_1^-  \chi _{c2}$  &  $3.28 \t 10^{-9}$ & $3.88 \t 10^{-9}$  \\ \cline{2-4}

& $B_c^- \to D_1^0  D_2^-$  &  $4.23 \t 10^{-7}$ & $2.36 \t 10^{-6}$ \\ \cline{2-4}

& $B_c^- \to \underline{D}_1^0  D_2^-$  &  $1.33 \t 10^{-7}$ & $7.43 \t 10^{-7}$ \\ \hline

\multirow{7}{*}{$\Delta S =  0$}
& $B_c^- \to D_{s1}^-  \chi _{c2}$  &  $3.34 \t 10^{-5}$ &  $3.95 \t 10^{-5}$ \\ \cline{2-4}
& $B_c^- \to \underline{D}_{s1}^-  \chi _{c2}$  &  $7.49 \t 10^{-6}$ &  $8.85 \t 10^{-6}$ \\ \cline{2-4}
& $B_c^- \to \chi _{\text{c1}}  D_{s2}^-$  &  $4.78 \t 10^{-7}$ &  $2.67 \t 10^{-6}$ \\ \cline{2-4}

& $B_c^- \to K_1^-  \bar{D}_2^0$  &  $2.46 \t 10^{-8}$  $(2.53 \t 10^{-8})$ $^{@}$ & $2.90 \t 10^{-8}$ $(2.81 \t 10^{-8})$ $^{@}$\\ \cline{2-4}

& $B_c^- \to \underline{K}_1^-  \bar{D}_2^0$  &  $3.66 \t 10^{-8}$ $(7.61 \t 10^{-9})$ $^{@}$  &  $4.33 \t 10^{-8}$ $(8.85 \t 10^{-9})$ $^{@}$\\  \cline{2-4}

& $B_c^- \to a_1^0  D_{s2}^-$  &  $4.55 \t 10^{-10}$ &  $2.54 \t 10^{-9}$ \\ \cline{2-4}

& $B_c^- \to f_1  D_{s2}^-$  &  $4.54 \t 10^{-10} $  &  $2.54 \t 10^{-9}$ \\
 \hline \hline

\end{tabular}
\begin{tablenotes}
      \centering \footnotesize
           \item $^{@}$  for  $\theta_K=-58^{\circ}$.
    \end{tablenotes}
\end{table}

\begin{table}
\captionof{table} {Branching ratios of $B_{c}   \to   AT$ decays for CKM-suppressed modes.}
\label{t4}
\begin{tabular}{c c l l} \hline \hline
\multirow{2}{*}{Mode}  &  \multirow{2}{*}{Decays} & \multicolumn{2}{c}{Branching Ratios} \\ \cline{3-4}
 \multicolumn{2}{c} {}& $N_c =  3$ & $N_c \to  \infty$ \\  \hline
\multirow{4}{*}{$\Delta S =  -1$}
& $B_c^- \to K_1^-  \chi _{c2}$  &  $1.66 \t 10^{-5}$ ($1.63 \t 10^{-5}$)$^{@}$&   $1.97 \t 10^{-5}$ ($1.90 \t 10^{-5}$)$^{@}$\\ \cline{2-4}

& $B_c^- \to \underline{K}_1^-  \chi _{c2}$  &  $2.40 \t 10^{-5}$ ($4.76 \t 10^{-5}$)$^{@}$& $2.84 \t 10^{-5}$ ($5.58 \t 10^{-5}$)  $^{@}$\\ \cline{2-4}

& $B_c^- \to D_1^0  D_{s2}^-$  &  $4.45 \t 10^{-8}$ &   $2.49 \t 10^{-7}$\\ \cline{2-4}

& $B_c^- \to \underline{D}_1^0  D_{s2}^-$  &  $1.40 \t 10^{-8}$ & $7.81 \t 10^{-8}$   \\ \hline

\multirow{7}{*}{$\Delta S =  0$}

& $B_c^- \to D_1^-  \chi _{c2}$  &  $2.32 \t 10^{-6}$ &   $2.74 \t 10^{-6}$\\ \cline{2-4}

& $B_c^- \to \underline{D}_1^-  \chi _{c2}$  &  $7.42 \t 10^{-7}$ & $8.77 \t 10^{-7}$  \\ \cline{2-4}

& $B_c^- \to \chi _{\text{c1}}  D_2^-$  &  $2.42 \t 10^{-8}$ & $1.35 \t 10^{-7}$  \\ \cline{2-4}

& $B_c^- \to a_1^-  \bar{D}_2^0$  &  $6.82 \t 10^{-7}$ &  $8.07 \t 10^{-7}$ \\ \cline{2-4}

& $B_c^- \to f_1  D_2^-$  &  $3.96 \t 10^{-9}$ & $2.21 \t 10^{-8}$  \\ \cline{2-4}

& $B_c^- \to a_1^0  D_2^-$  &  $3.96 \t 10^{-9}$ & $2.21 \t 10^{-8}$  \\ \cline{2-4}

& $B_c^- \to b_1^-  \bar{D}_2^0$  &  $4.77 \t 10^{-12}$ & $5.65 \t 10^{-12}$  \\  \hline \hline
\end{tabular}
\begin{tablenotes}
   \centering    \footnotesize
           \item $^{@}$ for   $\theta_K=-58^{\circ}$.
    \end{tablenotes}
\end{table}

\begin{table}
\captionof{table} {Branching ratios of $B_{c}   \to   AT$ decays for CKM-doubly-suppressed modes.}
\label{t5}
\begin{tabular}{c c l l} \hline \hline
\multirow{2}{*}{Mode}  &  \multirow{2}{*}{Decays} & \multicolumn{2}{c}{Branching Ratios} \\ \cline{3-4}
 \multicolumn{2}{c} {}& $N_c =  3$ & $N_c \to  \infty$ \\  \hline
\multirow{4}{*}{$\Delta S =  -1$}
& $B_c^- \to D_{s1}^-  \bar{D}_2^0$  &  $2.53 \t 10^{-7}$ & $3.00\t 10^{-7}$ \\ \cline{2-4}
& $B_c^- \to \underline{D}_{s1}^-  \bar{D}_2^0$  &  $8.43 \t 10^{-8}$ & $9.96 \t 10^{-8}$ \\ \cline{2-4}
& $B_c^- \to \bar{D}_1^0  D_{s2}^-$  &  $6.76 \t 10^{-9}$ & $3.78 \t 10^{-8}$ \\ \cline{2-4}

& $B_c^- \to \underline{\bar{D}}_1^0  D_{s2}^-$  &  $2.12 \t 10^{-9}$ & $1.19 \t 10^{-8}$ \\  \hline

\multirow{4}{*}{$\Delta S = 0$}

& $B_c^- \to D_1^-  \bar{D}_2^0$  &  $1.51 \t 10^{-8}$ & $1.79 \t 10^{-8}$ \\ \cline{2-4}

& $B_c^- \to \underline{D}_1^-  \bar{D}_2^0$  &  $4.75 \t 10^{-9}$ & $5.61 \t 10^{-9}$ \\ \cline{2-4}

& $B_c^- \to \bar{D}_1^0  D_2^-$  &  $1.75 \t 10^{-10}$ & $9.76 \t 10^{-10}$ \\ \cline{2-4}

& $B_c^- \to \underline{\bar{D}}_1^0  D_2^-$  &  $5.49 \t 10^{-11}$ & $3.07 \t 10^{-10}$ \\ \hline \hline

\end{tabular}
\end{table}

\end{document}